\documentclass[preprint2]{aastex}
\usepackage{graphicx}
\usepackage{natbib}

\shorttitle{New M dwarf debris disk candidates in NGC 2547}
\shortauthors{Forbrich et al.}

\begin{document}


\title{New M-dwarf debris disk candidates in NGC~2547}


\author{Jan Forbrich, Charles J. Lada, August~A. Muench, \& Paula S. Teixeira\altaffilmark{1,2}}
\affil{Harvard-Smithsonian Center for Astrophysics, 60 Garden Street, Cambridge, MA 02138}
\email{jforbrich, clada, gmuench, pteixeira@cfa.harvard.edu}
\altaffiltext{1}{also at: Departamento de F\'{\i}sica da Faculdade de Ci\^encias da \hbox{Universidade} de Lisboa, Ed. C8, Campo Grande, 1749-016, \hbox{Lisboa},Portugal}
\altaffiltext{2}{also at: Laborat\'orio Associado Instituto D. Luiz - SIM, Universidade de Lisboa, Campo Grande, 1749-016, \hbox{Lisboa}, Portugal}

%\and

%\author{R. J. Hanisch\altaffilmark{5}}
%\affil{Space Telescope Science Institute, Baltimore, MD 21218}


%% \altaffiltext{1}{Visiting Astronomer, Cerro Tololo Inter-American Observatory.
%% CTIO is operated by AURA, Inc.\ under contract to the National Science
%% Foundation.}
%% \altaffiltext{2}{Society of Fellows, Harvard University.}
%% \altaffiltext{3}{present address: Center for Astrophysics,
%%     60 Garden Street, Cambridge, MA 02138}
%% \altaffiltext{4}{Visiting Programmer, Space Telescope Science Institute}
%% \altaffiltext{5}{Patron, Alonso's Bar and Grill}


\begin{abstract}
With only six known examples, M-dwarf debris disks are rare, even though M dwarfs constitute the majority of stars in the Galaxy. After finding a new M dwarf debris disk in a shallow mid-infrared observation of NGC~2547, we present a considerably deeper \textsl{Spitzer}-MIPS image of the region, with a maximum exposure time of 15~minutes per pixel. Among sources selected from a previously published membership list, we identify nine new M dwarfs with excess emission at 24~$\mu$m tracing warm material close to the snow line of these stars, at orbital radii of less than 1~AU. We argue that these are likely debris disks, suggesting that planet formation is under way in these systems. Interestingly, the estimated excess fraction of M stars appears to be higher than that of G and K stars in our sample.
\end{abstract}





\section{Introduction}

Even though about 80~\% of the stars in the Galaxy are M~dwarfs \citep{lad06}, only very few of these are known to harbor debris disks. This is even more surprising since many M~dwarfs do show signs of protoplanetary disks in earlier stages of evolution with similar frequency as more massive stars (e.g., \citealp{and05}, \citealp{muz06}, \citealp{lad06b}) and we now know seven M dwarfs that harbor extrasolar planets (most recently \citealp{end08}, see references therein). While protoplanetary disks contain large amounts of primordial gas and dust out of which planets form, debris disks are older and entirely made of collisionally evolving planetesimals and dust particles. Debris disks, thus, are secondary products of the planet formation process and their existence in a given system implies previous formation of protoplanets. Compared to optically thick protoplanetary disks, debris disks are fainter and optically thin. For a recent observational and theoretical review of debris disks, see \citet{mey07} and \citet{mor07}. 

Since the cold dust in debris disks is best studied at far-infrared (FIR) or submillimeter wavelengths, the \textsl{Spitzer} Space Telescope, given its unprecedented FIR sensitivity, has revolutionized the research on the role of debris disks in star and planet formation. 
Most of the \textsl{Spitzer} studies of debris disks have been carried out using the Multiband Imaging Photometer for Spitzer \citep[MIPS,~][]{rie04}, operating at wavelengths of 24~$\mu$m, 70~$\mu$m, and 160~$\mu$m. 

Prior to our study, only six debris disks around M dwarfs were known. Early studies based on data obtained by IRAS, the Infrared Astronomical Satellite, reported far-infrared excess emission towards a small number of field M stars \citep{tsi88,mul89,mat91}, but few of these results could be confirmed subsequently \citep[e.g.,][]{son02,ria06}. The first firm detection of an excess was the disk of AU~Mic \citep{kal04,liu04b}, a member of the $\beta$~Pic moving group with an age of $\approx$12~Myr. The discovery observation was made with a stellar coronograph detecting scattered light in near-infrared broad-band filters. AU~Mic was also found to have excess emission at 70~$\mu$m, but not at 24~$\mu$m \citep{che05}. Subsequently, in submillimeter radio observations at 850~$\mu$m, \citet{liu04a} identified GJ~182, a member of the Local Association Group with an age of $\approx$50~Myr, as harboring a debris disk. GJ~182 does not show a far-infrared excess in the \textsl{Spitzer}-MIPS study of \citet{che05}. In \textsl{Spitzer}-MIPS data of NGC~2547, a cluster at an age of 30--40~Myr, \citet{you04} detected the third M~star debris disk (source 23 in our Table~\ref{tabsigexcl}, see also Teixeira et al., \textsl{in prep.}). TWA~7 was the fourth object reported; it is a member of the TW Hya association with an age of $<6$~Myr and excess emission reported at 24~$\mu$m, 70~$\mu$m, and submillimeter radio wavelengths \citep{low05,mat07}. One field M dwarf (GJ~842.2) was found to have a debris disk \citep[in submillimeter observations,][]{les06}. Finally, \citet{gor07} identified one additional M~dwarf debris disk in NGC~2547 at 24~$\mu$m, listed as source 22 in our Table~\ref{tabsigexcl}. 

Only three sources (TWA~7, source 22, and source 23) have been identified as far-infrared excess sources at 24~$\mu$m, while the other sources have been found at either 70~$\mu$m  (AU Mic) or submillimeter radio wavelengths. Since material of quite different distances from the central star is probed with these two techniques, debris disks identified in the mid-infrared and in submillimeter radio are not necessarily directly comparable to one another.

Studying a sample of 62 field M dwarfs estimated to be older than 1~Gyr, \citet{gau07} did not find any indication of excess emission at 24~$\mu$m or 70~$\mu$m indicative of debris disks (see also \citealp{pla05,ria06}). In contrast to these \textsl{Spitzer} results, \citet{les06} discuss two sets of (sub-)millimeter data, including previously published observations, and find three detections of excess emission in a sample of 23 M~dwarfs with ages ranging from 20--200~Myr, corresponding to an excess fraction of $13^{+6}_{-8}$~\%, which may not be significantly different from the disk fractions of earlier spectral types.

After identifying an M~dwarf debris disk in earlier \textsl{Spitzer} observations of NGC~2547 (source 23, see above), we performed ten times longer \textsl{Spitzer}-MIPS 24~$\mu$m observations of the same region in order to find out whether this was a singular object or part of a larger population. In the remainder of the introduction, we summarize the current knowledge about NGC~2547 before describing the observations in Section~\ref{sec_obs}. We present the results in Section~\ref{sec_results}, including remarks on the 70~$\mu$m and 160~$\mu$m data as well as variability at 24~$\mu$m, discuss these results in Section~\ref{sec_disc} and close with a summary in Section~\ref{sec_summ}.

\section{Background: Cluster Parameters}
NGC~2547 is a young open cluster at a distance of $433_{-48}^{+62}$~pc \citep{rob99} and an average visual extinction of only $A_V=0.18$~mag \citep{cla82}. It has an approximate age of $\sim30-40$~Myr \citep[][who fit four different evolutionary tracks to \textsl{UBVRI} photometry, three of them yielding the older age]{lyr06}. \citet{nay06} determine an age of $38.5_{+3.5}^{-6.5}$~Myr (and a lower distance of $361^{+19}_{-8}$~pc) from a fit to a color-magnitude diagram, improving on an earlier age estimate in the range of 20 -- 35~Myr \citep[][based on \textsl{BVI} photometry]{nay02}. 
\citet{you04} discuss the disk population based on \textsl{Spitzer} imaging data, including one epoch of the MIPS data presented in this paper, as does \citet{gor07}. Both of them use the data we mark as ``epoch 1''.

\begin{figure*}[t]
\centering
%~/ngc2547/mips24PRN_N.eps -> f1.eps
\epsscale{1.8}
\plotone{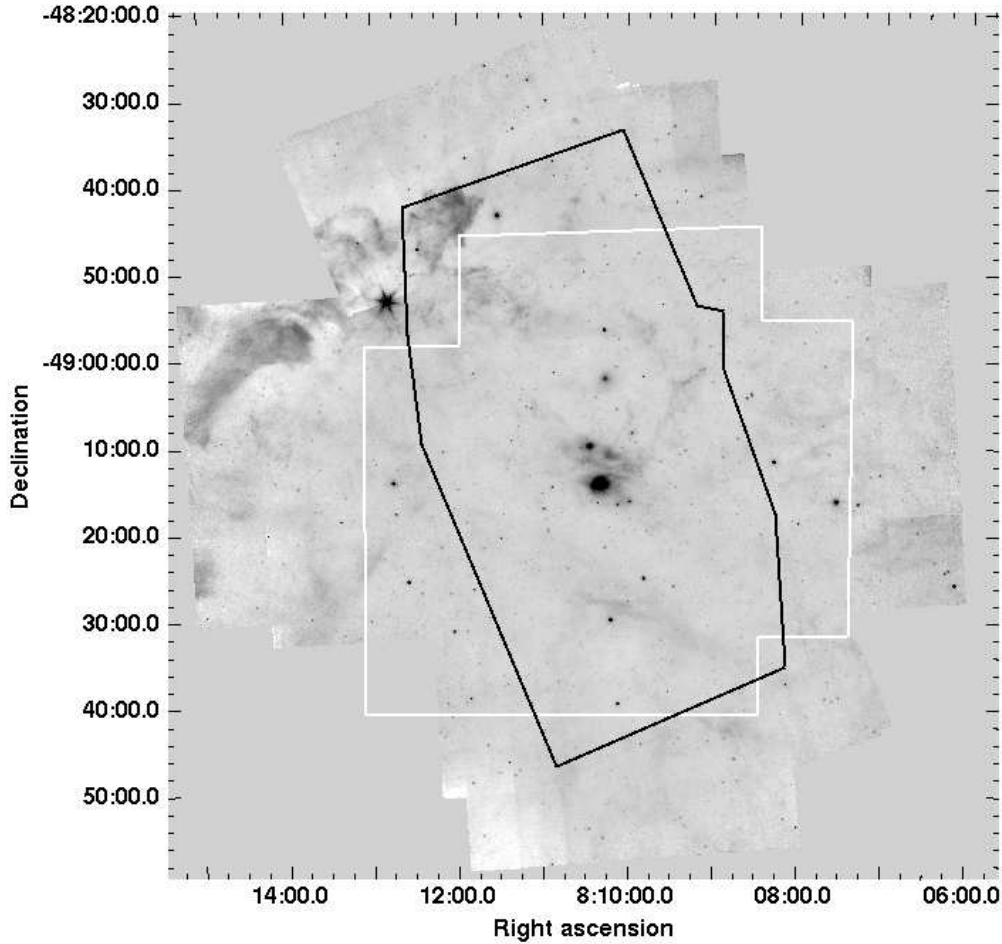}
\caption{MIPS-24~$\mu$m map of NGC~2547. The black line delineates the ``core area'' with at least two thirds of the maximum exposure while the white line contains the ``highly probable members'' as defined by \citet{gor07}. The image scale is logarithmic and the pixel size is 2.45$''$. \label{mips24PRN}}
\epsscale{1.0}
\end{figure*}

\citet{gor07} provide a membership list based on a deep \textsl{RIZ} survey
carried out by \citet{jef04}. While the latter paper distinguishes empirical candidate members derived using different evolutionary tracks, \citet{gor07} use photometric, spectroscopic, and proper motion information for different subsamples to produce a single list of 806 ``highly probable members''. However, 64 of these sources appear not to warrant such a classification because they are located in the two fields that \citet{jef04} used to check for galactic contamination, far from the actual cluster (outside the coverage of Fig.~\ref{mips24PRN}). Therefore, the total number of ``highly probable members'' is 742. Twelve sources in the full list do not have 2MASS-\textsl{K} magnitudes, although they are detected in $J$ band. \citet{gor07} also provide near-infrared spectra for 89 sources. \citet{irw08} determine rotation periods at optical wavelengths for 176 photometrically selected candidate members. 128 of these have counterparts in the ``highly probable members'' listed by \citet{gor07}.

\section{Observations and Data Reduction}
\label{sec_obs}

We present a deep \textsl{Spitzer}-MIPS observation of the central part of NGC~2547. The combined dataset that we are discussing here consists of all MIPS maps taken of NGC~2547 to date; it includes the one map analyzed by \citet{gor07}. We took ten additional maps, but due to an instrument failure during observation, one map was lost. Thus, we discuss a total of ten different maps that can be grouped into three epochs (Table~\ref{tbl-obs}). All but epoch 1 also have 70~$\mu$m data. Additional 160~$\mu$m data was taken in the three observations of epoch 2.

The entire 24~$\mu$m dataset consists of 21127 ``basic-calibrated data'' (BCD) files, processed with pipeline S14.4.0. These data were reduced using the \textsl{Spitzer} MOsaicker and Point source EXtractor (MOPEX, \citealp{mak05}), version 030106. In order to produce a single deep map out of single maps with different background levels, the MOPEX task \textsl{overlap.pl} was used on subsets of the full dataset after flatfield correction. The subsequent data reduction steps including median filtering, outlier detection and masking before coadding and combining were carried out on these subsets of BCDs. The resulting map is shown in Fig.~\ref{mips24PRN}. The map shows that the region is largely devoid of extended emission, as faint extended features can only be seen in the logarithmically stretched image. Subsequently, the \textsl{Spitzer} Astronomical Point Source EXtraction (APEX) software was used for photometry on the resulting maps, applying point-response-function (PRF) fitting with local noise estimation and a cutoff based on the signal-to-noise ratio (SNR). Prior to the actual photometry, a sample PRF was determined from several well-detected stars in the field. A number of objects were detected in several of the overlap-corrected map chunks which themselves are overlapping. In these cases, photometry was taken from the map with the higher exposure map value in order to ensure optimal coverage. The photometric accuracy was checked by subtracting the fitted sources from the original maps in order to create residual maps. Only two sources have residuals that are clearly off (even negative); both are bright sources within extended emission (see also discussion below). Corresponding steps were taken for the analysis of the 70~$\mu$m and 160~$\mu$m data. For comparison, point-response function photometry was also performed on the newly reduced archival IRAC data.

\begin{table}
\begin{center}
\caption{NGC~2547 MIPS 24~$\mu$m observations \label{tbl-obs}}
\begin{tabular}{lll}
\tableline\tableline
Epoch & UT start date \& time & further obs.\\
\tableline
1  & 2004-01-28 19:17:57.6  & --\\ % map00
2a & 2006-05-08 23:01:46.6  & 70,\,160\,$\mu$m\\ % map01
2b & 2006-05-09 07:29:42.3  & 70,\,160\,$\mu$m\\ % map10
2c & 2006-05-09 12:02:06.3  & 70,\,160\,$\mu$m\\ % map09
3a & 2006-12-01 22:01:58.8  & 70\,$\mu$m\\ % map02
3b & 2006-12-02 01:03:24.0  & 70\,$\mu$m\\ % map03
3c & 2006-12-03 21:36:33.4  & 70\,$\mu$m\\ % map06
3d & 2006-12-04 00:38:00.7  & 70\,$\mu$m\\ % map07
3e & 2006-12-04 03:44:48.9  & 70\,$\mu$m\\ % map04
3f & 2006-12-04 06:46:16.2  & 70\,$\mu$m\\ % map05 
\tableline
\end{tabular}
\end{center}
\end{table}

\section{Results}
\label{sec_results}

\subsection{24~$\mu$m data}
\subsubsection{Source detection}

At a 3$\sigma$ detection level, a total of 27042 sources is detected. This is our full sample with all detections in the full coverage area, including outlying areas which were covered only once. The effective exposure time per pixel in this full dataset varies by a factor of ten across the map. In order to have a more homogeneous dataset, we only use detections with at least 2/3 of the maximum effective exposure time per pixel ($t_{max}=901.6$~s) in a second sample. This second sample, which we will refer to as the core area, focuses on the central region of NGC~2547; it includes 6390 sources. The distribution of 24~$\mu$m magnitudes in this core area peaks at a value of 11.6~mag which compares to a peak at 10.8~mag in the epoch~1 data used by \citet{gor07}. Every single map roughly has 10\% of the total maximum effective exposure time per pixel.

\begin{figure}
\centering
\plotone{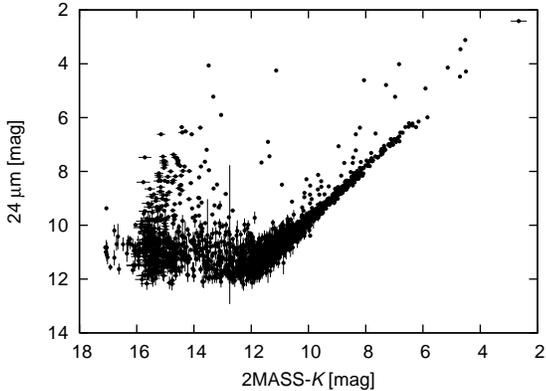}
\caption{MIPS-24~$\mu$m vs. \textsl{K} band (2MASS) photometry for the 1945 MIPS sources with 2MASS counterparts; error bars are nominal 1$\sigma$. A few sources have large error bars and some of the faintest 2MASS sources shown do not have photometric errors. \label{crossmatch_2mass_mips24e_1.5asec_1.5asec_24vsK}}
\end{figure}

\begin{figure}
\centering
\plotone{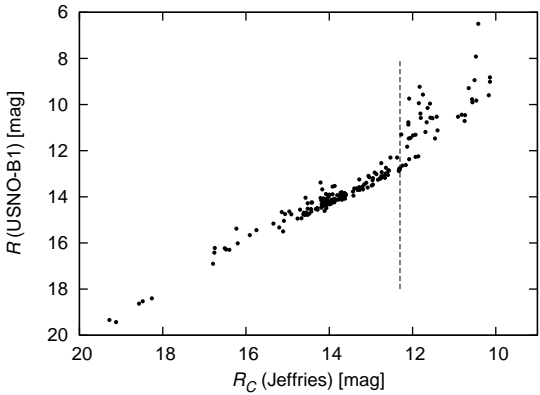}
\caption{Comparison of the \textsl{R$_C$} band magnitudes from \citet{jef04} with \textsl{R} band magnitudes from USNO-B1 for the 208 highly probable members detected at 24~$\mu$m in the core area. For sources brighter than 12.3~mag (dashed line), saturation effects become important, and we use the USNO \textsl{R} photometry instead of the \citet{jef04} photometry. \label{RjefvsRusno}}
\end{figure}

\begin{figure}[h]
\centering
\plotone{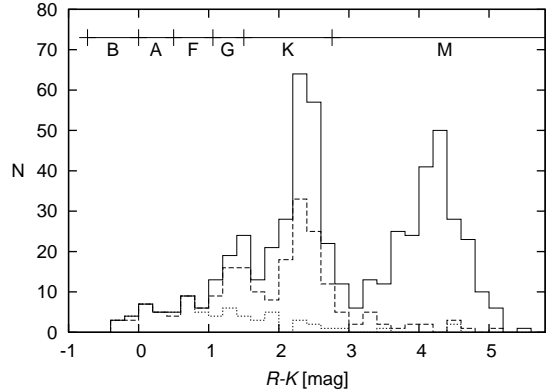}
\caption{Estimate of completeness in the core area. The solid-line histogram shows the \textsl{R-K} distribution of the 554 candidate members located in our core area (and having \textsl{R-K} photometry), the dashed histogram shows our 208 detections at 24~$\mu$m, and the dotted histogram shows the 25 detections at 24~$\mu$m reported by \citet{gor07}. The indicated spectral types from B stars to about M6 are based on \textsl{R-K} and are for luminosity class V, using intrinsic colors from \citet{bes88}, \citet{bes91}, and \citet{ken95} with updates from \citet{win97}. \label{R-Khist.eps}}
\end{figure}

As a first step, the 24~$\mu$m detections were correlated with the 2MASS point source catalogue since all candidate members should be NIR-detected. 
There are 1945 sources from the full list that have 2MASS counterparts within $1\farcs5$. This radius corresponds to 3 times the half-width half-maximum of a Gaussian distribution fitted to the angular separations between the 2MASS and the MIPS data; it was used for all subsequent correlations. 
Fig.~\ref{crossmatch_2mass_mips24e_1.5asec_1.5asec_24vsK} shows that there is a good correlation between 24~$\mu$m and \textsl{K} band magnitudes down to \textsl{K}$\approx12$. The scatter of 24~$\mu$m magnitudes for sources with a given \textsl{K} magnitude of $K>12$~mag indicates that these are likely galaxies instead of single stars since 2MASS photometric errors in $K$ increase significantly only beyond $K=14$~mag.

In fact, only 646 sources out of the 6390 sources in the core area have 2MASS counterparts, indicating that the overwhelming majority of the sources we detect at 24~$\mu$m are galaxies that are too faint in the \textsl{K} band to be detected by 2MASS. Out of the 6390 sources, 208 have counterparts in the Gorlova membership table. Correlating all 27042 sources in our entire dataset to the Gorlova list yields 265 counterparts. This means that the white polygon in Fig.~\ref{mips24PRN} encompasses 265 detections while the black polygon contains 208 detections. \citet{gor07} extracted 74 candidate members with 24~$\mu$m counterparts (from the map we list as Epoch 1); 64 of these are detected in the core area although it actually covers 71 of them. The seven missing sources either did not pass our SNR selection criterion or the angular distance between a detection and the catalog position was too large. The median magnitude difference at 24~$\mu$m between \citet{gor07} and our dataset is 0.04~mag, including five cases where the difference is larger than 0.5~mag (these are the bright sources within extended emission; the maximum difference is 1.5~mag).

The following discussion focuses on the core area. For the near-infrared photometry, we make use of the 2MASS and IRAC photometry from \citet{gor07}. The deepest optical data come from the $R_cI_cZ$ survey carried out by \citet{jef04}. We therefore focus on the \textsl{R-K} color baseline. 

Fig.~\ref{RjefvsRusno} shows a direct comparison of photographic \textsl{R} band magnitudes from the USNO-B1 catalog with those from \citet{jef04}. Because of $R_c$ band saturation problems for the brightest sources in this sample ($R_c<12.3$~mag), we use \textsl{R} magnitudes from the USNO-B1 catalog \citep{mon03} instead for those bright sources. The median difference in \textsl{R} between USNO-B1 and \citet{jef04} within the 208 candidate members is 0.13~mag, but it is considerably higher for the brightest sources (up to 3.9~mag). The photometric accuracy of the USNO-B1 catalog is given as 0.3~mag. By combining these two sets of photometry measurements, we can make use of reliable \textsl{R} band photometry for all 208 sources in our core sample, at the price of varying accuracy. For comparison, the \textsl{R} band photometry used by \citet{gor07} does not include information for 47 out of the 208 sources. 

There are a few exceptions to our $R<12.3$~mag selection rule. Due to a binary source fitted as a single source in USNO-B1, two sources with $R<12.3$~mag only have \citet{jef04} photometry which we therefore use as best choice. There are nine other sources in the sample which have unreliable USNO-B1 photometry for the same reason. Also, two obviously faulty \textsl{R} band magnitudes in the core sample ($R>28$~mag) were replaced by the USNO values. In total, we make use of USNO-B1 photometry for 53 out of the 208 sources.

In Fig.~\ref{R-Khist.eps}, we show the distributions in \textsl{R-K} of the 562 ``highly probable members'' with \textsl{R-K} photometry in our core area (using \citealp{jef04} and USNO-B1 photometry), our 208 detections at 24~$\mu$m, as well as those detections within our core area that were reported by \citet{gor07}. These authors report \textsl{R-K} for only 26 out of their 74 detections at 24~$\mu$m. For our analysis, we use the same combination of USNO-B1 and \citet{jef04} photometry for these sources that we also use for our own data to provide \textsl{R} magnitudes: seventy-one of the 24~$\mu$m sources from \citet{gor07} are located in our core area and seventy now have reliable \textsl{R} band magnitudes. While by far not detecting all candidate members, we detect a considerably larger fraction than \citet{gor07}. Fig.~\ref{R-Khist.eps} indicates a completeness dropping to $\sim$50\% in the K~stars. However, although not complete, we still get a good sampling of K and M stars. It becomes immediately clear that more spectroscopic confirmations of photometric members are needed since there is a lot of contamination, especially due to the giant branch at \textsl{R-K}$\approx$2.5. In fact, \citet{jef04} identify the region with 0.5$<$\textsl{R-I}$<$0.7 as particularly prone to contamination. There are 167 out of 562 candidate members in our core region that fall into this range which corresponds to a full range in \textsl{R-K} of 1.84 to 2.68. However, as will become clear later, this does not affect the discussion of the M stars.

Looking at the entire dataset, only 120 out of 265 highly probable members detected at 24~$\mu$m have available \textsl{V-K} colors, ranging from --0.36 to 5.73. In the core area, this ratio is only 111 out of 208. Since, with the above procedure, we have \textsl{R-K} for all 208 sources of interest, we used \textsl{R-K} colors instead of \textsl{V-K} as a proxy for spectral type.

\subsubsection{Excess sources}
\label{subsec_excess}

\begin{figure*}[]
\epsscale{2.10}
\plotone{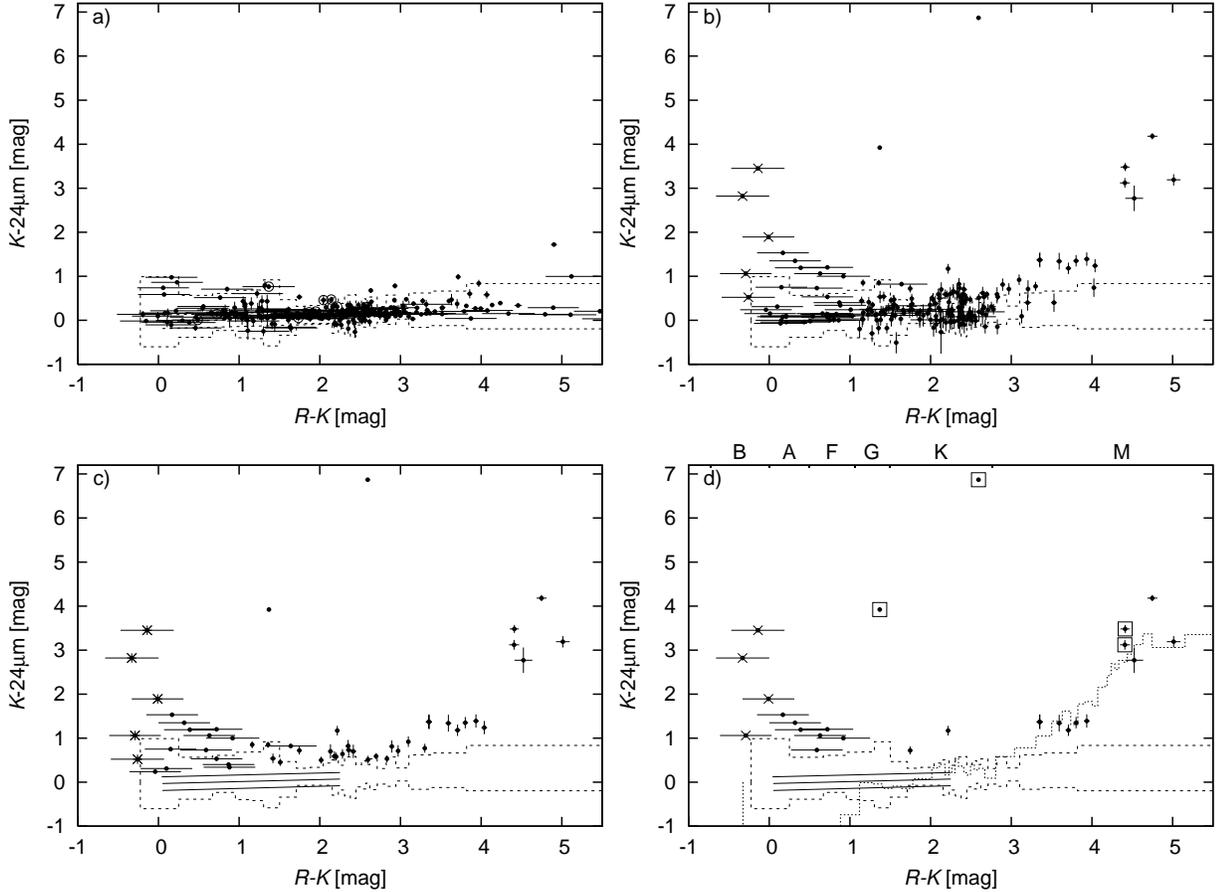}
\caption{Plots of \textsl{K}-24~$\mu$m vs. $R-K$ for a) 275 24~$\mu$m-detected sources in the core area and not in the ``highly probable member'' list (14 matches with the ``less plausible'' members from \citet{gor07} are marked by `$\circ$', but only a few are slight outliers), b) 208 24~$\mu$m-detected highly probable members in the core area, c) the 54 significant excess sources according to the Pleiades criterion, and d) the 24 excess sources outside the color-color exclusion zone; boxes denote the four significant excess sources (using our exclusion zone criterion) identified by \citet{gor07} in the core area. The solid lines indicate the locus of the Pleiades and respective 3$\sigma$ limits, as used by \citet{gor07}, and the dashed lines delineate the 3$\sigma$ color-color exclusion zone (see text). A measure of completeness, the dotted line shows the mean of the distribution of 554 ``highly probable members'' sources with \textsl{R} and \textsl{K} magnitudes in the core area \citep{gor07}, assuming a limiting 24~$\mu$m magnitude of 11.6 mag. Sources with large error bars in $R-K$ have \textsl{R} band photometry from USNO-1B (see text). The `$\times$' symbol in panels b), c), and d) marks five sources that are located in bright extended emission and were already designated as ``halo-like'' by \citet{gor07}. The error bars show the propagated $1\sigma$ errors. The spectral types indicated in panel d), from B stars to about M6, are based on \textsl{R-K} and are for luminosity class V. \label{K-24vsR-K_plots}}
\epsscale{1.0}
\end{figure*}

We used two different criteria to identify sources with excess emission at 24~$\mu$m. Initially, following \citet{gor07}, we used the 3$\sigma$ locus of the Pleiades stars in a \textsl{K}-24 vs. \textsl{R-K} diagram of our data. The Pleiades locus was derived by \citet{gor06,gor07}.
However, since the membership list for NGC~2547 is not yet definitive, we also looked at the distibution of all sources detected at 24~$\mu$m in the same area for which \textsl{R} and \textsl{K} photometry is available, including 14 ``less plausible'' members from \citet{gor07}. To crudely exclude extragalactic sources, we only use candidate non-members with $K<12$~mag (see above), yielding 275 sources in the core area. We characterize their location in the color-color diagram by determining the standard deviation in 25 \textsl{R-K} bins of equal sample size, clipping the most extreme point in every bin. The resulting color-color exclusion zone is defined as the 3$\sigma$ variation in every bin. Note that for the \textsl{R}-band photometry, we use the same combination of USNO-B1 and \citet{jef04} photometry that was also used for the members (see above).

According to these criteria, 54 sources have significant excess emission compared to the Pleiades, i.e., their 3$\sigma$ photometric error bars lie completely outside the 3$\sigma$ locus of the Pleiades. Twenty-six of these sources also are outside the color-color exclusion zone (including sources embedded in bright extended emission). In the following, we use the latter as the more conservative criterion to define excess sources. The results of the search for excess sources are summarized in Fig.~\ref{K-24vsR-K_plots} and in Table~\ref{tabsigexcl}. Panel a) of Fig.~\ref{K-24vsR-K_plots} shows the definition of the color-color exclusion zone by plotting all 275 candidate non-member sources within the core area having 24~$\mu$m detections as well as \textsl{R} and \textsl{K} magnitudes (with $K<12$). Panel b) shows all 24~$\mu$m--detected ``highly probable members'' \citep{gor07} within our core area, with the near-infrared photometry explained above. The lower panels show the significant excess sources when using the Pleiades (panel c) and the color-color exclusion zone (panel d) criteria. Both Pleiades locus and exclusion zone are shown in all plots. Panel d) additionally shows the previously identified four excess sources within the core area (using the exclusion zone criterion) \citep{gor07} as circles. A fifth excess source in \citet{gor07}, at \textsl{R-K}=1.95~mag and \textsl{K}-24=1.91~mag, is not detected in our data (and not shown in the plot) because the positional difference between the 24~$\mu$m source and the NIR counterpart is too large ($3.6''$ in our case). Since the 24~$\mu$m magnitudes are nearly identical, this is probably the same source, however.

\begin{figure}
\centering
\epsscale{1.0}
\plotone{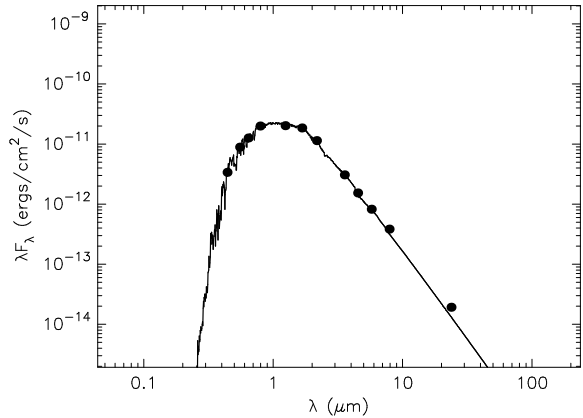}
\caption{SED of 2MASS~J08095644-4922109, a confirmed M0 member of NGC~2547 \citep[source 37 in][]{jef03} \textsl{without} significant 24~$\mu$m excess emission. The fit uses the Phoenix model \citep{hau99a,hau99b}, with log(g)=5 and solar metallicity; the derived temperature is T=4100~K. \label{m0spec}}
\end{figure}

We estimate sensitivity across \textsl{R-K} by assuming a limiting magnitude at 24~$\mu$m and calculating the distribution in \textsl{K-24~$\mu$m} vs. \textsl{R-K} of the 554 ``highly probable members'' located in the core area that have suitable photometry (eight sources lack 2MASS-\textsl{K} photometry). We then determine the standard deviation in \textsl{K}-24~$\mu$m for 46 bins containing nearly the same sample sizes, with a fixed value at 24~$\mu$m of 11.6~mag (see above), thus characterizing the distribution per \textsl{R-K} bin.
Also this analysis takes into account USNO-B1 photometry for the brightest sources. Panel d) of Fig.~\ref{K-24vsR-K_plots} shows the mean of this distribution. It becomes clear that for late M dwarfs, we can \textsl{only} detect excess sources, whereas for earlier spectral types, we can also detect photospheres, although the detections of K and M dwarfs are not complete. 
In fact, most of the M dwarfs that we detect at 24~$\mu$m (14 out of 25) do not show significant excess emission. To corroborate this result, we show the SED of a spectroscopically confirmed M0 member that we detect as a source with a significance of 4.6~$\sigma$ at 24~$\mu$m at presumably photospheric emission levels in Fig.~\ref{m0spec}. 

\citet{gor07} report 26 sources with 24~$\mu$m detections as well as \textsl{R} and \textsl{K} photometry, of which nine have excess emission compared to the Pleiades and six have excess emission with respect to our color-color exclusion zone. Only one of these sources is not in our core area. Another source from within these six sources is 3.6$''$ from the nearest 2MASS source in our data, so we do not list it. The remaining four previously identified excess sources are marked as circles in panel d) of Fig.~\ref{K-24vsR-K_plots}. 

As remarked by \citet{gor07}, five of the 24~$\mu$m-brightest sources covered in both their data and our core area are embedded in bright extended emission. Four out of these five technically have excess emission, even according to the color-color exclusion zone criterion, but the extended structures are too large to represent debris disks. 
We compared the the PRF-fitting photometry with aperture photometry and found that the the former appears to reasonably reflect the actual source brightness in the central areas while the curve of growth of the latter does not converge. The photometric results for these bright sources with extended emission have to be interpreted with caution.

\begin{figure}
\centering
\plotone{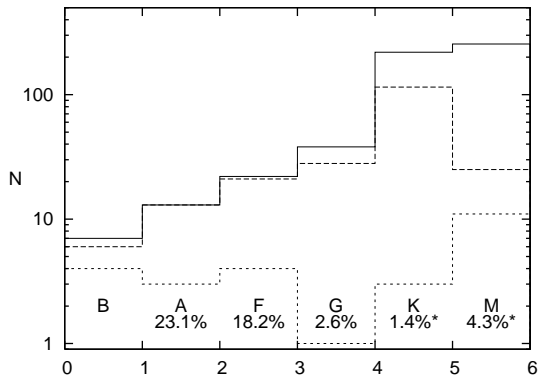}
\caption{MIPS-24~$\mu$m excess sources as a function of spectral type. The uppermost histogram shows all 554 ``highly probable members'' with \textsl{R-K} inside our core area, the next histogram shows the 208 sources we detect at 24~$\mu$m, while the lowest histogram shows the 26 significant excess sources outside of the color-color exclusion zone. The excess fractions are given as a percentage of the 554 candidate members in the area. No percentage is given for the B stars because some of them are excess sources with extended emission that clearly are not debris disks (see text). The excess fraction for K and M~stars are marked with an asterisk because the data probably are not complete for these spectral types (see text). \label{det_exc_comp_spec}}
\end{figure}

Even though we are dealing with relatively low numbers, it is interesting to look at the excess fraction per spectral type, as derived from \textsl{R-K} colors, which we plot in Fig.~\ref{det_exc_comp_spec}. While the contamination of the membership list makes it difficult to state exact fractions, this problem is mostly constrained to the K~stars (see Fig.~\ref{R-Khist.eps}), where many of the ``highly probable members'' may be background giants. Even though our data are not complete for M~dwarfs, 11 of 26 significant excess sources are M dwarfs, corresponding to a lower limit of 4.9\% of all ``highly probable'' member M stars in the area, or 44\% of the M dwarfs we detect at 24~$\mu$m. Four of these 11 M~dwarfs are spectroscopically confirmed members (Table~\ref{tabsigexcl}).
The excess rate is is of particular interest given the minimum in the G stars, where we detect just a single significant excess source.

Warmer protoplanetary disks closer to the central star would be recognizable by their mid-infrared excess emission at shorter wavelengths, as probed by the \textsl{Spitzer}-IRAC bands. Referring otherwise to the discussion of the available IRAC data in \citet{you04} and \citet{gor07}, who found an excess fraction of $<$1\% in the IRAC bands, we note in particular that there is virtually no sign of excess emission in the IRAC bands towards the M dwarfs with 24~$\mu$m excess emission (eight out of eleven are covered by the IRAC data; see also the discussion of the SEDs). The only excess detection is source 23, listed as ID~7 in \citet{gor07}, which has an 8$\sigma$ excess at 8$\mu$m.

\begin{figure}
\centering
%\epsscale{.46}
\epsscale{0.8}
\plotone{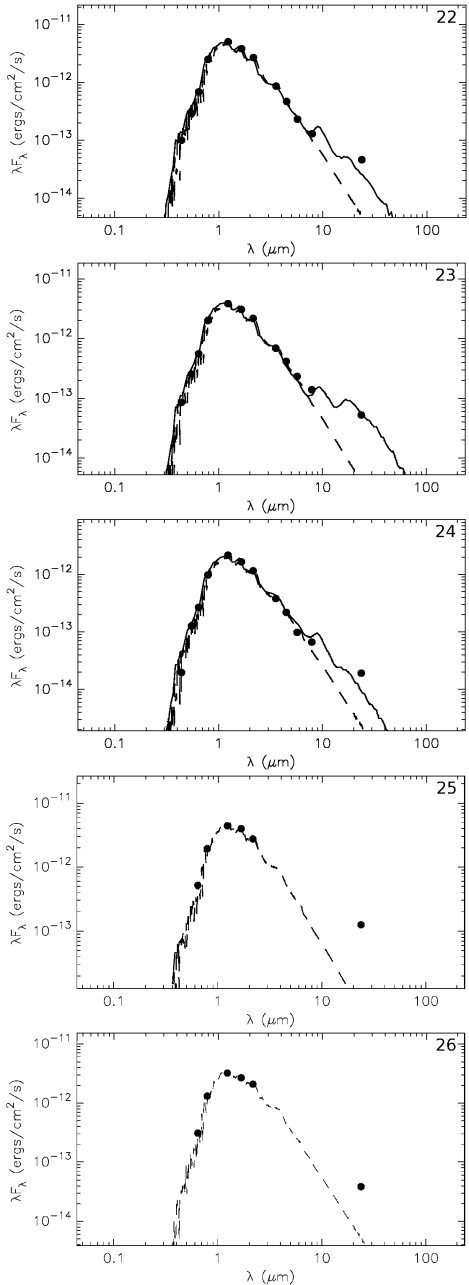}
\caption{SEDs fitted with disk models from \citet[][solid line]{rob07} and the  Phoenix atmosphere model (dashed line) for the five excess sources with \textsl{R-K}$>4$. Two sources are located outside the footprint of the IRAC mosaic, for them, only the atmosphere model is shown. Source numbers are indicated in the upper right corner of every panel. \label{sedplots}}
\end{figure}

\begin{deluxetable}{rllllllllllr}
\tabletypesize{\scriptsize}
%\rotate
\tablecaption{MIPS-24~$\mu$m excess sources outside the color-color exclusion zone, sorted in \textsl{R-K}\label{tabsigexcl}}
%\tablewidth{0pt}
\tablecolumns{12}
\tablehead{
\colhead{no.} & \colhead{2MASS} & \colhead{\textsl{K}} & \colhead{\textsl{R-K}} & \colhead{err.} &
\colhead{\textsl{K}-24} & \colhead{err.} & \colhead{$24\mu$m} & \colhead{err.} & \colhead{spec?\tablenotemark{a}} &
\colhead{comments\tablenotemark{b}}& \colhead{SNR}\\
\colhead{} & \colhead{} & \colhead{[mag]} & \colhead{[mag]} & \colhead{[mag]} & \colhead{[mag]} & \colhead{[mag]} & \colhead{[mag]} & \colhead{[mag]} & \colhead{} & \colhead{}& \colhead{$24\mu$m}
}
\startdata
 1 & J08102058--4914144&   6.83 &  -0.33&   0.33&   2.82&    0.03&    4.01&  0.00&    & B3III/IV, X, G, extd  &   17.1\\   % jef98 RX 59 (r<2"), HD 68478 B3III/IV
 2 & J08101607--4902058&    7.65&  -0.29&   0.32&   1.06&    0.02&     6.59&  0.00&    & B5II, X, G, extd &   54.0\\  % jef98 RX 56 (r<3"), B5II
 3 & J08102725--4909509&   8.06 &  -0.14&   0.33&   3.45&    0.03&    4.61&  0.00&    & B6V, X, G, extd   &  119.9\\   % HD68496 B6V
 4 & J08095753--4908202&   8.95 &  -0.01&   0.32&   1.89&    0.02&    7.06&  0.00&  2 & A0Vn, G, extd	     &   89.2\\   % HD68396 A0Vn
 5 & J08100607--4914180&   9.66 &   0.17&   0.32&   1.53&    0.03&    8.13&  0.01&  2 & A1V		     &   17.8\\   % NSV17775 A1V, cataclysmic
 6 & J08092668--4914371&   10.14&   0.32&   0.32&   1.35&    0.03&    8.79&  0.01&  2 &  		     &   46.2\\   %
 7 & J08100841--4900434&   9.55 &   0.39&   0.32&   1.19&    0.03&    8.37&  0.01&    & A3V		     &   77.0\\   % HD68420, A3V
 8 & J08111134--4904442&   10.18&   0.59&   0.32&   0.73&    0.04&     9.46&  0.02&    & 		     &   12.4\\   %
 9 & J08084571--4923473&   9.95 &   0.63&   0.32&   1.06&    0.03&    8.89&  0.01&    &  		     &   42.5\\ %
10 & J08110323--4900374&   10.15&   0.72&   0.32&   1.20&    0.03&    8.96&  0.01&    &  		     &   53.0\\   %
11 & J08093815--4918403&    9.67&   0.92&   0.33&   1.00&    0.04&    8.67&  0.01&  2 & A9V, G~ID\,2, X	&   39.2\\   % A9V
12 & J08090250--4858172&   11.36&   1.37&   0.02&   3.92&    0.02&    7.44&  0.00&  2 & G~ID\,8	     &  162.1\\   %
13 & J08101799--4923502&   11.65&   1.75&   0.03&   0.72&    0.09&    10.93&  0.07&  2 & X		     &    9.2\\   % jef98 RX 58 (r<2")
14 & J08102955--4922094&   12.35&   2.22&   0.03&   1.17&    0.11&    11.18&  0.09&    &		      &    6.0\\   %
15 & J08101691--4856291&   11.12&   2.59&   0.03&   6.87&    0.02&    4.25&  0.00&  2 & G~ID\,9, 70~$\mu$m	      &   1500\\   %
16 & J08095330--4918132&   12.85&   3.35&   0.04&   1.37&    0.15&    11.48&  0.12&  4 & M2, RV, X        &    3.9\\   % jef98 RX 33 (r<4"), M2 (jef03), u.l. for Li
17 & J08110007--4904413&   13.05&   3.35&   0.04&   1.37&    0.17&    11.68&  0.14&    &		      &    3.3\\   %
18 & J08101374--4917112&   13.16&   3.59&   0.04&   1.34&    0.19&    11.82&  0.16&    & X, P	      &    3.1\\   %
19 & J08105210--4921138&   12.52&   3.71&   0.04&   1.18&    0.13&    11.34&  0.10&    & X  	      &    6.5\\   %
20 & J08092450--4928575&   12.68&   3.80&   0.04&   1.35&    0.13&    11.33&  0.10&    & P			      &    5.5\\   %
21 & J08094510--4846520&   12.82&   3.94&   0.04&   1.39&    0.15&    11.43&  0.12&    &			      &    5.9\\   % 
22 & J08091770--4908344&   13.85&   4.41&   0.06&   3.12&    0.11&    10.73&  0.06& 1,3& M4.5, RV, P, G 	      &    6.8\\   % NO Li in \citet{oli03}, Src 41, Li<0.17A (2sigma), \citet{jef05}, r<1"<5": Src 1, Li<0.03A, confirmed member; M4.5
23 & J08093547--4913033&   14.07&   4.41&   0.06&   3.48&    0.09&    10.59&  0.05& 1,2,3& M4.5, RV, G~ID~7\tablenotemark{c}    &	8.5\\	% NO Li in \citet{oli03}, Src 50, Li<0.16A (2sigma), \citet{jef05}, r<1"<5": Src 7, Li<0.04A, confirmed member
24 & J08085407--4921046&   14.76&   4.52&   0.11&   2.77&    0.29&    11.99&  0.20&  1 &		      &    4.8\\ % NO Li in \citet{oli03}, Src 82, Li<0.33A (2sigma)
25 & J08104437--4939001&   13.82&   4.75&   0.06&   4.18&    0.07&    9.64&  0.02&  3 & M5, RV  	      &   25.2\\   % Jef04 ID 774, Status 1 RV member
26 & J08101391--4939481&   14.11&   5.01&   0.09&   3.19&    0.13&    10.92&  0.07&  1\tablenotemark{d}     &  &    7.7\\ %Li in \citet{oli03}, Src 67, LiEW=0.555±0.27A !!!
%FOUR Gorlova "halo" sources would have sign. excess, even using the exclusion zone, ONE additional compared to Pleiades (122.748149  -49.284424)
%ONE Gorlova source is mismatch (3.98"): 122.603803 -48.863407
\enddata
%% Text for table notes should follow after the \enddata but before
%% the \end{deluxetable}. Make sure there is at least one \tablenotemark
%% in the table for each \tablenotetext.
%\tablecomments{Table \ref{tbl-1} is published in its entirety in the 
%electronic edition of the {\it Astrophysical Journal}.  A portion is 
%shown here for guidance regarding its form and content.}
\tablenotetext{a}{Near-infrared spectrum in 1) \citet{oli03}, 2) \citet{gor07}, 3) \citet{jef05}, 4) \citet{jef03}.}
\tablenotetext{b}{spectral type from SIMBAD, `X' = XMM detection within $<2.5''$, from \citet{jef06},`P' = optical period from \citet{irw08}, `G' = previously identified 24~$\mu$m source \citep{gor07}, `RV' = confirmed rad. vel. member \citep{jef03,jef05}, `extd' = in bright extended emission, possibly unreliable 24~$\mu$m photometry}
\tablenotetext{c}{discovered by \citet{you04}}
\tablenotetext{d}{{Lithium} absorption with EW$_{\rm Li}=0.56\pm0.27$~\AA}
\end{deluxetable}

\subsection{70~$\mu$m and 160~$\mu$m data}

After data reduction, we performed PRF-fitting photometry on the 70~$\mu$m and 160~$\mu$m data using APEX at a 5$\sigma$ detection limit. 
In an image built from the filtered 70~$\mu$m BCD data, 390 sources are detected. Within search radii of $2''$ and $4''$, only one of the ``highly probable members'' is clearly detected, with an SNR of 15; this is source 15, also showing the strongest excess emission at 24~$\mu$m, its 70~$\mu$m magnitude is 2.76~mag.
\citet{gor07} concluded that it may be an equal-mass binary late K dwarf (their source ID~9).
For all other sources, upper limits from aperture photometry were obtained, but they do not constrain the spectral energy distributions of these sources and thus are not discussed further.

At 160~$\mu$m, only the extended emission features are recognizable, otherwise none of the ``highly probable members'' are detected. Also here, upper limits derived from aperture photometry do not constrain the source SEDs.

\subsection{Variability}

The fact that the 24~$\mu$m observations were carried out in three distinct epochs offers the possibility to study variability on timescales of months. Looking for large variability, PRF-fitting photometry was run separately on all ten maps. Then, by cross-correlation, for every source the number of detections in these ten maps was determined. As a (rather crude) measure of variability, the standard deviation of the set of photometry measurements was calculated for every source.

At a 3$\sigma$ detection level, 38 highly probable members are detected in all ten 24~$\mu$m maps. As a measure of variability, we determine for every source the ratio of the maximum (presumed) variability amplitude divided by the noise level. When merging the photometry into one value for each epoch, only four out of 38 sources have a variability exceeding the respective 5$\sigma$ level, and only two sources exceed 7$\sigma$. These are 2MASS J08101546-4905487, one of the brightest sources in the field, varying by 19\%, and 2MASS J08085649-4923128, varying by 24\%; both sources are not among the excess sources discussed above.

\section{Discussion}
\label{sec_disc}

There are three different possibilities for the interpretation of the observed excess emission \citep[e.g.,][]{stru06}. It could be due to remnant primordial material, due to a debris disk with continuously replenished dust, or it could be debris produced by a recent catastrophic collision of planetesimals, i.e., debris not in (quasi-) equilibrium. 
The dust in debris disks is produced and replenished by collisions, and then removed by processes such as the Poynting-Robertson (PR) drag, radiation pressure and stellar winds. Direct radiation pressure does not play a role for the low-luminosity stars that we consider \citep{pla05}. The balance of dust production and these removal processes determines the amount of observable dust.

%===> Which radius?
In order to discuss the physical processes, we first need to determine which region around a central star is probed by the 24~$\mu$m (excess) emission. As an estimate, we consider for a given stellar luminosity the orbital radius of a blackbody which is at a temperature that puts the peak of its SED at 24~$\mu$m. For a dwarf star with a luminosity of $L$=0.01~$L_\odot$, the corresponding radius is only 0.5~AU. 

%===> PR drag and timescale -> debris disk
With the orbital radius of the dust particles constrained, we can now determine the lifetime of dust particles under the influence of the PR drag, which should yield an upper limit of the dust removal timescale \citep[e.g.,][]{dom03}. The PR lifetime of a dust particle in orbit around a central star scales with particle size and density as well as the square of its orbital radius and is inversely proportional to the luminosity of the central star \citep{bur79}. For micron-sized particles with a density of $\rho=2.5$~g\,cm$^{-3}$ \citep{div93,che01} located at a distance of 1~AU from a central star with a luminosity of $L=0.01\,L_\odot$, the lifetime is still on the order of $1.7\times10^5$ years. This value is ten times longer for 10~$\mu$m-sized particles.  Since the M~dwarfs discussed here have an age of 30--40~Myr, the particles would need to be replenished and thus cannot be primordial in order to still be observable at that age, indicating the debris disk nature of these systems and ruling out the first possibility mentioned in the beginning of this section. The lifetimes do not approach the cluster age until an orbital radius of 13~AU. Since other dust removal processes like stellar winds potentially operate at the same time, the derived timescale is in fact an upper limit.

This result thus suggests that at an age of 30--40~Myr, M~dwarfs can have warm inner debris disks. Given that we know already seven M dwarfs that harbor extrasolar planets orbiting close to their host stars, this finding reinforces the idea that planets form around M stars.  With M~star debris disks more frequent than previously thought, chance occurrences like catastrophic collisions of big planetesimals may not have to be invoked for their explanation, ruling out the remaining third possibility.

%===> Remarks on radius
It is interesting to note that a blackbody with a peak of its SED at 24~$\mu$m has a temperature of 120~K, not far from the temperature of water sublimation, 153~K, defining the ``snow line'' \citep{hay81}, separating the inner region of rocky planet formation from the outer region of icy planet formation and playing a role in the definition of ``habitable zones''. 
\footnote{We can estimate a permissable range of dust temperatures for the dust we detect at 24\,$\mu$m. Comparing the ratio of an assumed 3$\sigma$ excess at 8\,$\mu$m and the observed 24\,$\mu$m excess with that expected for blackbody emission suggests an approximate upper limit of $\sim260$~K for a typical source without an 8\,$\mu$m excess. As a strict lower limit, we note that the M-dwarf debris disk around AU~Mic which does not exhibit 24\,$\mu$m excess emission has a dust temperature of $\sim40$~K \citep{ria06}.}
For M stars, the habitable zone lies at orbital radius of about 0.1~AU \citep{sca07}. Indeed, seven M dwarfs are now known to harbor one or several extrasolar planets at orbital radii of 0.02 to 2.3~AU (most recently \citealp{end08}, see references therein), with masses spanning 0.016--2.1~$M_J$. The five planets with the smallest semi-major axes ($<0.1$~AU) have masses of $<0.07$~$M_J$. During the early evolution of an M dwarf, the snow line moves inwards due to the decreasing luminosity of the central star, enabling the formation of icy protoplanets within the first 1~Myr at orbital radii of 1--4~AU, not taking into account subsequent migration \citep{ken07}. The snow lines of M~dwarfs in NGC~2547 have already moved further in when compared to their initial location, so that rocky protoplanets may have previously formed at orbital radii larger than the current snow line.

%===> Comp to submm
Given that some of the few previously known M dwarf debris disks have been found at submillimeter wavelengths, it is important to keep in mind that material traced in the far infrared is quite different from material traced in the submillimeter radio range. The latter corresponds to material at temperatures of $<20$~K at orbital radii of several hundred AU \citep[e.g., ][]{les06}. Thus, different regions are probed by far-infrared and submillimeter observations and they cannot be directly compared.

%===> SEDs
We show the photometric spectral energy distributions of the five M~dwarfs with the strongest excess emission at 24~$\mu$m in Fig.~\ref{sedplots}. Their 24~$\mu$m emission is about an order of magnitude stronger than the expected photospheric flux. To obtain approximate physical properties from the crudely constrained photometric SEDs, we performed fits using disk models by \citet{rob07}. 
For one of our best-constrained disks, source 23, the fits suggest that slightly more than 1\% of the total luminosity is due to the disk, even though the central object is a star with a luminosity of only $L=0.028$~$L_\odot$. 
To within an order of magnitude, the resulting dust disk masses are $\sim10^{-9}$~$M_\odot$,
corresponding to only a few percent of one lunar mass, but in some cases are barely constrained.
As a major source of uncertainty, these mass estimates scale with the assumed dust properties which would be much better constrained with (sub-) millimeter data. For all but the densest regions of the disk, the model assumes small dust particles slightly larger than those in the diffuse ISM, and therefore, the mass estimates are a lower limit.
Reviewing the submillimeter detections of M~dwarf debris disks, \citet{les06} quote masses of one to 13 lunar masses for three debris disks in the M0 range. In marked contrast, \citet{che05} use \textsl{Spitzer}-MIPS data to derive a minimum dust mass for the debris disk of AU~Mic of only 10$^{-4}$ lunar masses, assuming a particle size of 0.2~$\mu$m. This mass difference may indicate that submillimeter and far-infrared--detected debris disks are indeed quite different. However, a comparison of both methods towards the same objects would be needed to confirm this conjecture.

%===> age effect?
All of the M~dwarf members of NGC~2547 can be assumed to be of the cluster age, 30--40~Myr, possibly indicating that at this age, the evolutionary stage of M~dwarfs makes it easier to detect such debris disks. 
For debris disks around stars of earlier spectral types, there have already been several studies concerning age trends. In a study of main-sequence A stars older than 20~Myr, \citet{rie05} find that the magnitude of excess at 24~$\mu$m declines with a $t_0/t$ dependence, with $t_0\approx 150$~Myr. Studying the double cluster h and $\chi$ Persei in context with this result and other observations, \citet{cur08a} conclude that the debris disk excess emission of A~stars peaks at about the age of this cluster, 13~Myr. On longer timescales, the evolution is less clear. While \citet{tri08} did not find a clear age trend studying FIR excess emission of solar-type field stars with ages ranging from 200~Myr to 10~Gyr, \citet{mey08}, also studying solar-type stars, found an excess rate decreasing from 18\% at ages 3--30~Myr to 2\% at ages 0.3--3~Gyr. \citet{hil08} find that the 70~$\mu$m excess emission of 328 FGK stars, tracing cooler dust compared to 24~$\mu$m, peaks at ages of 30--100~Myr.

For a solar-mass star, the debris for rocky planet formation peaks after a few million years for material at 1~AU \citep{ken04,ken05}. To scale this result for an M dwarf, we can use a simple relation for the disk mass, $M_{\rm disk} \propto M_{\rm star}$, and the collision timescale, $t_{\rm coll} \propto M_{\rm star}^{-1/2}$ \citep{ken08}. So, if the M dwarf is three times less massive than the Sun, we would expect planet formation to take about five times longer around an M dwarf, i.e., 10--20 Myr, not taking into account minor complications due to the moving snow line. This indicates that the M dwarfs in NGC~2547 are beyond but still close to the maximum predicted debris production. Especially with regard to stars of earlier spectral types, it is important to keep in mind that this argument only allows to compare the evolutionary timescale of debris at an orbital radius of 1~AU. For A stars, the radius probed by 24~$\mu$m observations is beyond 20~AU, and the timescales at these radii may be very different.

%===> known constraints on evolution
Currently, the early debris-disk evolution of M dwarfs is poorly constrained from an observational point of view. \citet{wei04} did not find significant mid-infrared excess (12 and 18~$\mu$m) towards 16 members of the young, nearby TW Hydrae association and conclude that apparently any planet formation in the terrestrial planet region was rapidly completed. In a study targeting stars at age similar to the age of NGC~2547, \citet{mam04} observed members of the 30-Myr-old Tucana-Horologium Association, including two M dwarfs, at a wavelength of $\sim$10~$\mu$m (\textsl{N} band). They do not find excess emission towards the two M dwarfs, and their quoted uncertainties would allow excess emission at only $\sim$10\% of the expected photospheric flux level. Looking at our SEDs in Fig.~\ref{sedplots}, if the excess emission is modeled correctly, it would be far more difficult to detect at 10$\mu$m.
The M dwarfs studied by \citet{pla05}, and not found to exhibit excess emission at 11.7~$\mu$m, include one young source and nine more objects with ages of $>$600~Myr. The single young star is GJ~3305, a member of the $\beta$~Pic moving group at an age of $\sim$12~Myr. 
\citet{gau07} found that field M~dwarfs older than 1~Gyr do not show far-infrared excess emission. Due to the nearby location of these sources ($d<20$~pc), the absolute sensitivity of these observations is much higher than that of our dataset. This upper limit for the evolutionary timescale of M~dwarf debris disks indicates that planet formation around M dwarfs has ceased by this time.

%===> earlier spectral types
We finally consider the excess fractions of stars with earlier spectral types. While \citet{gor07}, restricting the discussion to their completeness limit, found 30\ -- 45\% of the B--F members to show excess emission at 24~$\mu$m, we interestingly find virtually no G~star excess emission at this wavelength. As pointed out above, while we find excess emission towards K stars, we may not be complete throughout their \textsl{R-K} range, and many of the ``highly probable'' K members may be background giants. Even so, the 24~$\mu$m excess rate for both G and K stars, using our conservative criterion for excess sources, is at most a few percent. In the most comprehensive sample of field stars to date, \citet{tri08} find statistically indistinguishable excess rates for A, F, G, and K stars, with  an average age of the solar-type sample of 5~Gyr. The fact that we do not find excess emission towards G~stars can possibly be understood using an argument similar to the one used in our discussion of M dwarf debris disks: given the cluster age, the solar-type stars are far beyond their peak in dust production. In marked contrast, the excess rate for M dwarfs is surprisingly high, even though it is a lower limit for two different reasons: first, our 24~$\mu$m detections of M dwarfs are not complete and second, we use a conservative criterion to define excess sources. It remains unclear, however, to what degree the excess rates of G, K, and M stars are skewed due to contamination by non-members.

\section{Summary}
\label{sec_summ}

In an exceptionally deep 24~$\mu$m observation of NGC~2547, we detect eleven M~dwarf excess sources interpreted as debris disks, nine of which are new. These constitute more than 42\% of all significant excess sources that we find across the range from B to M stars. The eight M dwarf excess sources that were covered by IRAC observations do not show significant excess in these mid-infrared bands (except for source 23, see \citealp{gor07}).
The dust masses of these disks are crudely estimated to be on the order of a few percent of a lunar mass. The identification as debris disks hinges on the observation that the dust removal timescale due to processes like PR drag and stellar winds is much smaller than the age of the cluster. Thus, there must be a mechanism to replenish the dust particle population, e.g., by collisions. Given the cluster age, the observed M dwarfs appear to be close to, maybe beyond their maximum predicted debris production. Interestingly, the observed debris material is located close to the snow line and the habitable zone, at less than 1~AU from the central stars. Since even our deep dataset is not complete for M~dwarfs, we cannot give a debris disk fraction in terms of the total population, but it appears to be higher than for G and K stars at the same age. In fact, we find a minimum in excess emission in the G stars, suggesting that their maximum debris production occurs much earlier than at 30--40~Myr.

With these new discoveries, the number of known M~star debris disks more than doubles by rising from six to fifteen. The previously perceived ``dearth'' of M dwarf debris disks apparently was due to small sample sizes both in numbers of objects studied and in different ages probed. To better constrain the debris-disk evolution and planet formation timescales of M dwarfs, larger samples of objects at ages both younger and older than NGC~2547 are needed.

\acknowledgments{We would like to thank Scott Kenyon for insightful discussion and comments. Observations reported here were obtained through NASA \textsl{Spitzer} GO program (PID 20124) and supported by JPL contract 1281114. P.~S.~T. acknowledges support from the scholarship SFRH/BD/13984/ 2003 awarded by the Funda\c{c}\~ao para a Ci\^encia e Tecnologia (Portugal). }


%% in parentheses, after the keyword, but they will not be verified.

{\it Facilities:} \facility{SST (MIPS)}

\end{document}